# The VQR, Italy's second national research assessment: methodological failures and ranking distortions[1]

**Giovanni Abramo***

*Laboratory for Studies of Research and Technology Transfer. Institute for System Analysis and Computer Science - National Research Council of Italy (IASI-CNR) Viale Manzoni 30, 00185 Rome, Italy - giovanni.abramo@uniroma2.it*

**Ciriaco Andrea D'Angelo**

*University of Rome "Tor Vergata", Dept of Engineering and Management Via del Politecnico 1, 00133 Rome, Italy - dangelo@dii.uniroma2.it*

**Abstract**

The 2004-2010 VQR, completed in July 2013, was Italy's second national research assessment exercise. The VQR performance evaluation followed a pattern also seen in other nations, in being based on a selected subset of products. In this work we identify the exercise's methodological weaknesses and measure the distortions that result from them in the university performance rankings. First we create a scenario in which we assume the efficient selection of the products to be submitted by the universities and from this simulate a set of rankings applying the precise VQR rating criteria. Next we compare these "VQR rankings" with those that would derive from application of more appropriate bibliometrics. Finally we extend the comparison to university rankings based on the entire scientific production for the period, as indexed in the Web of Science.

*\* Corresponding author*

# 1. Introduction

It is widely held that higher education and research institutions play crucial roles in socio-economic development and competitiveness, particularly in the era of the knowledge economy. For many governments, research and higher education policies are thus near the top of the policy agenda. The education systems of various nations naturally assume different forms, however analysis of national policies still shows certain common traits, such as an emphasis on continuous improvement, internationalization and the pursuit of excellence (Stoker, 2006). To support these aims, there is a growing trend to implement national exercises for the evaluation of research activity in universities and public institutions. The specific objectives of the exercises vary but include: i) provision of guidance for merit-based allocation of public funding; ii) continuous improvement in research productivity through comparative analysis of performance and subsequent selective funding; iii) identification of strengths and weaknesses in disciplines and geographic areas, to support formulation of research policy at governmental level and management strategies at institutional levels; iv) provision of convincing information to taxpayers on the effectiveness of research management and delivery of public benefits, and; v) reduction of information asymmetry between new knowledge users and suppliers. In the current authors' view, the last of these objectives is the most important, subject to the important provision that the comparative evaluation of performance extends to the level of the individual researchers. With this provision, individual users (students, enterprises and funding agencies) are informed of the research productivity of individual scientists and of their grouping in schools and universities and can make the most effective choice of new knowledge producers for their needs. This in turn has an incentive effect on the new knowledge producers to improve their offer. Proper comparative evaluation thus has the potential to improve the efficiency of the knowledge market and initiate continuous improvement processes in the entire education system[2].

Given the national objectives at stake, administrators and policy makers face very important and delicate duties in providing correct formulation of research assessment exercises and equally challenging duties in the proper communication and use of the results. Errors at any stage can make the whole process dysfunctional. In particular, errors in performance ranking lists or mistaken communication of procedures and progress may jeopardize the achievement of overall objectives.

There are currently 15 nations (China, Australia, New Zealand, 12 EU countries) that conduct regular comparative performance evaluations of universities and link the results to public financing (Hicks 2012). The shares of overall public funding and the criteria for assigning funds vary from nation to nation. However the selective funding policies in place seem prone to risks and critical problems. Geuna and Martin (2003) analyze practices in twelve European and Asia-Pacific nations: they show that while initial benefits may outweigh the costs, selective funding seems to produce diminishing returns over time. Severe criticism has been mounted concerning the methodology of the UK assessment exercises (Adams and MacLeod, 2002; Sastry and Bekhradnia, 2006; Lipsett, 2007; Martin and Whitley, 2010), as well those of the Australian (Butler, 2003a, 2003b; Steele et al., 2006) and Italian evaluations (Abramo et al., 2013a; Abramo and D'Angelo, 2011).

---

[2] In Italy the lack of performance assessment at the individual level does not permit an efficient selection: in 95% of cases, private companies activate their research collaborations with academics less qualified than others and in 65% of cases these collaborations are also with academics at universities further away than those having more qualified experts (Abramo et al., 2011).

Given the crucial expectations for research assessment exercises, scholars specializing in this particular should normally be called on for their input: i) ex ante, for the definition of the project in a manner that follows state-of-the-art methodologies, coherent with the policy makers' macroeconomic objectives; ii) post facto, for determination of the results with respect to the declared objectives.

In Italy, the Ministry of Education, University and Research (MIUR) recently brought the latest national evaluation exercise to a close. The MIUR entrusted implementation of the national exercise to the newly formed Agency for the Evaluation of University and Research Systems (ANVUR), which opened the evaluation process at the end of 2011. The 2004-2010 VQR (Research Quality Evaluation) terminated on 16 July 2013, with the publication of the university performance ranking lists. The results will now determine allocation of an important share of financing for the individual institutions. The exercise was based on a hybrid peer-review/bibliometrics approach[3], in which universities were required to submit, for each of their professors, the best three research products from the 2004-2010 period. The choice of three products by ANVUR seems to have been dictated by the convergence of budget constraints and legislated impositions, requiring that the quality of research products be evaluated mainly by peer-review methodology[4]. As a corollary, one of the practical effects of the legislated regulation was to render it impossible to conduct bibliometric assessment of all professors' products in the hard sciences.

In this work we provide a critical examination of the VQR evaluation methodology for the hard sciences in light of the most recent literature on bibliometric techniques. We intend to apply our analyses in stages, in a manner that distinguishes the effects arising from the ANVUR choices and those arising from legislated regulation. For this, we first view the assessment of three products as a fixed constraint, and ask whether the bibliometric evaluation envisaged by ANVUR could have been better. Subsequently, we remove the constraint of three products and extend the assessment to all individuals' products. The analytical methodology is to empirically measure the consequences of the VQR's divergences (ANVUR/legislative origins) from more appropriate and feasible assessment design in terms of their impact on the university performance scores. Since the actual products submitted by the universities are not revealed, our comparison requires assumptions in the assessment: i) that the products the universities could present were limited to those indexed in the Web of Science (WoS); ii) that from these, the universities selected the best products (efficient selection).

The actual comparisons proceed in three steps:
1) For each professor in the hard sciences, we extract the three best products from the WoS and assign to each of these a value according to the precise terms defined under the VQR. Then draw up the resulting performance ranking lists of the universities (see Section 5).
2) We repeat the exercise but now selecting and rating the best three products per professor on the basis of more appropriate bibliometric criteria, and then compare the resulting performance scores to the "VQR scores"[5] from point 1).

---

[3] http://www.anvur.org/sites/anvur-miur/files/bando_vqr_def_07_11.pdf, last accessed on March 21, 2014
[4] ANVUR was instituted by national DPR law No. 76 of 01/02/2010. This law also stated that the quality of research products is to be assessed mainly by peer review.
[5] For simplicity we term the ranking list based on simulation of the precise VQR criteria as the "VQR

3) We conduct a second comparison of the VQR ranking list to a list developed using bibliometric criteria but no longer limited to the three research products per staff member, as per the terms of the VQR. Instead the university rankings are based on the entire 2004-2010 scientific production from all the scientists, as indexed in the WoS.

The aim of the article is to identify and discuss the methodological failures of a national research assessment exercise, in order that in future they can be avoided by all countries engaging in similar exercises.

In the next section of the paper we summarize the critical issues in methodology concerning national research assessment exercises. Sections 3 and 4 provide a brief description of the Italian VQR, and in particular the criteria for scoring products submitted for evaluation in the hard sciences. Section 5 provides the comparison between the results from the first two simulations above, while Section 6 gives the comparison between the first and third simulations. The work concludes with a summary of the principal results and the authors' considerations.

## 2. National research assessment exercises: critical issues and their impact

The need for national research evaluation is fully agreed at the theoretical level, but issues are more problematic when it comes to what methods to apply. The recent development of bibliometric techniques has led various governments to introduce bibliometrics in cases where they can be applied in support or substitution of more traditional peer review. In the UK, the 2014 Research Excellence Framework (REF) will be an informed peer-review exercise, with the assessment outcomes being a product of expert review informed by citation information and other quantitative indicators. The REF substitutes the previous Research Assessment Exercises, which were pure peer review. The Excellence in Research for Australia initiative (ERA), in 2010, was conducted through a pure bibliometric approach for the hard sciences. Single research outputs were evaluated by a citation index referring to world and Australian benchmarks. The Italian 2004-2010 VQR is a "mixed" type of evaluation exercise, based primarily on bibliometric analysis for some disciplines but on peer review for others. This is a change from the first exercise (2001-2003 VTR), which was entirely based on peer review.

The pros and cons of peer review and bibliometric methods have been thoroughly dissected in the literature (Horrobin, 1990; Moxham and Anderson, 1992; MacRoberts and MacRoberts, 1996; Moed, 2002; van Raan, 2005; Pendlebury, 2009; Abramo and D'Angelo, 2011). The literature does not indicate decisively whether one method is better than the other but demonstrates that, for evaluation of individual scientific products, there is certainly a positive correlation between peer-review results and citation indicators (Aksnes and Taxt 2004; Oppenheim and Norris 2003; Rinia et al., 1998; Oppenheim 1997; van Raan, 2006). There is also positive correlation between peer-review ranking and bibliometric ranking conducted at the individual level (Meho and Sonnenwald, 2000) and at the level of organizations (Thomas and Watkins, 1998; Franceschet and Costantini, 2011; Abramo et al., 2011). In terms of accuracy,

---

list", in spite of the fact that the rankings differ from those of the real list. The real VQR is based on a selection of products that is not necessarily efficient and is not restricted to those publications indexed in the WoS.

robustness, validity, functionality, time and costs, the superiority of bibliometrics compared to peer review emerges when it is applied to large-scale comparative evaluation of departments or entire institutions (Abramo and D'Angelo, 2011). Abramo et al. (2010a) specifically examine the case of limiting the evaluation to a subset of the total organizational scientific production, which is an unavoidable aspect of all peer-review based exercises, for reasons of time and cost. They demonstrate that this aspect is a critical concern for the accuracy and robustness of the performance scores. In fact, as noted above, the UK REF will assess the work of only selected research units, and from these only a selection of their outputs. Similarly, in the Italian VQR, each university was asked to present the best three research products achieved by each professor over the period 2004-2010. In contrast, Australia's ERA, a bibliometric exercise for the hard sciences, requires the entire research staff of every institution to submit their full research product for evaluation. The latter approach offers at least two clear advantages:
- it avoids the distortion of performance due to inefficient selection of products for evaluation, on the part of individual scientists and of their institutions;
- it avoids distortions due to evaluating only a part of the research product.

Abramo et al. (2009) first quantified these distortions for the case of Italy's 2004-2006 VTR evaluation. For the recent VQR, Abramo et al. (2014) have again estimated the error in the selection of products for the hard sciences: the results indicate a worsening the maximum score achievable by 23% to 32%, compared to the score from an efficient selection. It is clear that whenever an evaluation is conducted by peer review, the exercise must necessarily be based on a subpopulation of products, for reasons of time and money. However if the evaluation exercise is based on bibliometric techniques and indicators this limitation no longer occurs. In fact the choice to limit the evaluation to a subpopulation is difficult to justify from cost or management points of view, even if the products chosen were consistently the excellent ones.

Given the above discussion, and the fact that other nations also base their evaluations on methodological choices similar to the VQR, we then ask: how reliable and robust are the resulting national ranking lists? To respond, we carry out several bibliometric simulations, as follows.

## 3. The 2004-2010 VQR Italian research assessment exercise

Until 2009, core government funding for Italian universities was input oriented: funds were distributed in a manner that would equally satisfy the needs of each and all, in function of institutional size and disciplines of research. The core funding, known as Ordinary Finance Funds (FFO) accounted for 56% of total university income. It was only following the first national evaluation exercise (2004-2006 VTR) that a minimal share, equivalent to 3.9% of total income, was allocated in function of research and teaching assessments. The launch of the second Italian evaluation exercise (VQR) was preceded by vigorous debate, further fueled by heavy cuts in financing to research and higher education under a series of governments. This debate saw opinions ranging from demands for courageous action from policy makers in planning and implementing a true performance-based research funding system to attain improved performance at all levels, to contrasting insistence on complete renunciation of any such initiative, or at least its serious revision. The VQR thus began in a period of heightened tensions. The purpose of the exercise was to evaluate research activity carried out over the 2004-2010 period as conducted by:
- state universities;
- legally-recognized non-state universities;
- research institutions under the responsibility of the MIUR[6].

The objects of evaluation are the institutions, their macro-disciplinary areas and departments, but not the individual researchers. The results influence two areas of future action: overall institutional evaluations will guide allocation of the merit-based share of FFO (13% in 2013 and increasing in subsequent years); evaluation of the macro-areas and departments can be used by the universities to guide internal allocation of the acquired resources.

The evaluation of the overall institutions is determined by the weighted sum of a number of indicators: 50% based on a score for the quality of the research products submitted and 50% derived from a composite of six other indicators (10% each for capacity to attract resources, mobility of research staff, internationalization and PhD programs; 5% each for ability to attract research funds and overall improvement from the previous VTR).

ANVUR nominated 14 evaluation panels (GEVs)[7] of national and foreign experts, one for each university disciplinary area (UDA) composing the national academic system. The institutions subject to evaluation were to submit a specific number of products for each person on their research staff, in function of academic rank and their period of activity over the seven years considered. The demand for university faculty members was for up to a maximum of three products, while for research institutions the maximum expectation was six products per person. ANVUR defined the acceptable products as: a) journal articles; b) books, book chapters and conference proceedings; c) critical reviews, commentaries, book translations; d) patents; e) prototypes, project plans, software, databases, exhibitions, works of art, compositions, and thematic papers.

Any results produced in collaboration with professors in the same institution could only be presented once. Thus professors were asked to identify a larger set of products than the minimal demand, from which the administration could then complete the selection of the number required for the VQR evaluation. The products were then submitted to the appropriate GEVs based on the professor's identification of the field for each product. The GEVs were to judge the merit of each product as one of four values:

A = Excellent (score 1), if the product places in the top 20% on "a scale of values shared by the international community";
B = Good (score 0.8), if the product places in the 60%-80% range;
C = Acceptable (score 0.5), if the product is in the 50%-60% range;
D = Limited (score 0), if the product is in the bottom 50%.

The institutions are also subject to potential penalties:
- in proven cases of plagiarism or fraud (score -2);
- for product types not admitted by the GEV, or lack of relevant documentation, or produced outside the 2004-2010 period (score -1);
- for failure to submit the requested number of products (-0.5 for each missing product).

---

[6] Other public and private organizations engaged in research could participate in the evaluation by request, subject to fees.
[7] Acronym of "Groups of Evaluation Experts"

## 4. The VQR criteria for scoring products in the hard sciences

In view of their discipline characteristics, each GEV restricted the types of products admissible from the broad list and further defined the criteria and methods of evaluation. GEVs 10 to 14 opted exclusively for peer review, while the hard science GEVs, numbered 1 to 9, chose a mixed evaluation approach consisting of:
- bibliometric analysis, for articles indexed in the two major international databases (WoS and Scopus);
- peer review, for all other products, or where requested by the subject institution for indexed articles that were the result of work in an emerging area, of interdisciplinary character, or highly specialized.

For products subject to bibliometric evaluation, the judgments (A, B, C, D) were determined by a combination of two different indicators:
- a first indicator, named $I_R$, linked to the impact factor of the publishing journal (for those indexed in the WoS) or to the SCImago Journal Rank (for those indexed in Scopus);
- a second indicator, named $I_C$, linked to the number of citations received by the article as of 31 December 2011.

Given the world distributions of these indicators for each subject category and year, the products were then assigned to four classes, applying both the journal impact factor and citation references:
- class 1, if the article (journal) places in the first quintile (top 80%-100%) of the citation (IF) reference world distribution;
- class 2, if the article (journal) places in the 60%-80% range;
- class 3, if the article (journal) places in the 50%-60% range;
- class 4, if the article (journal) places in below the median of the citation (IF) reference world distribution.

Each product was thus attributed to one of the 16 possible combinations of the four classes for each indicator. Each GEV defined its own algorithm for deriving the ultimate judgment scores (A, B, C, D) from these combinations. Some GEVs further differentiated their algorithms on the basis of the product age, meaning the year of publication. Table 1 and Table 2 present the example of the classification matrices for the Chemistry UDA. As shown in the Table 1 matrix, the Chemistry GEV decided that for mature products (articles published in the 2004-2008 period), more weight would be given to citations in determining the final merit judgment. For recent products (publications from 2009-2010), judgment was instead based primarily on the journal impact. The notation "IR" indicates situations where the values of the two indicators are inconsistent, with high citations but low values for journal impact or vice versa. In these cases, the GEV decided to submit the products to informed peer review, meaning that the reviewers provided a judgment "informed" by the values of the bibliometric indicators.

| $I_R \rightarrow$ $I_C \downarrow$ | 1 | 2 | 3 | 4 |
|---|---|---|---|---|
| 1 | A | A | A | IR |
| 2 | B | B | B | IR |
| 3 | IR | C | C | C |
| 4 | IR | D | D | D |

*Table 1: Classification matrix for 2004-2008 products in the Chemistry UDA; IR =*

| | | | | | | | | | | |
|---|---|---|---|---|---|---|---|---|---|---|
| *"evaluated by Informed Peer Review"* | | | | | | 3 | A | B | C | D |
| | | | | | | 4 | IR | IR | IR | D |

*Table 2: Classification matrix for 2009-2010 products in the Chemistry UDA. IR = "evaluated by Informed Peer Review"*

| $I_R \rightarrow$ $I_C \downarrow$ | 1 | 2 | 3 | 4 |
|---|---|---|---|---|
| 1 | A | IR | IR | IR |
| 2 | A | B | C | D |

The other GEVs defined algorithms for merit judgment that are similar to the Chemistry example. The notable variations are:

- the Mathematics and computer science, Physics and Agricultural and veterinary science GEVs used only one algorithm for products of all ages;
- the Biology and Medicine GEVs chose the WoS as the only reference to be used in bibliometric evaluation and collapsed some of the WoS subject categories into larger groupings, meaning that the reference distributions for both $I_R$ and $I_C$ are given by the merging of all world publications in the grouped categories[8];
- for $I_C$, in the Biology, Medicine, Earth science and Agricultural and veterinary science GEVs, the calculations for "articles" and "reviews" are based on distinct citation distributions;
- for $I_R$, the Mathematics and computer science and Industrial and information engineering GEVs drew on a combination of indicators from various sources and published a document with the resulting classification (1-4) of each journal; in particular, the Mathematics and computer science GEV established very stringent criteria for inclusion of journals in "class 1";
- the Agricultural and veterinary sciences GEV provided notice that review articles published in a pre-established list of journals would automatically be routed to evaluation by peer review.

All GEVs agreed on including self-citations for calculation of $I_C$.

It is difficult to identify any literature on evaluation that would support some of the specific criteria adopted under the VQR. Much of the evaluation design can ultimately be traced to ANVUR's decision to assign the GEVs the responsibility of defining the requirements and criteria. This would assume that the members of the GEVs had specialized competencies in the area of research of evaluation, which was not necessarily the case. Thus in the following section we conduct an analysis of the changes in the scores and shifts between the rankings obtained under the VQR-GEV criteria and an identical exercise where the scoring is based on a single indicator supported by current literature on bibliometric evaluation.

## 5. Simulating the VQR and testing the robustness of its scoring criteria

We now compare the results from two different evaluation scenarios. The first scenario is based on the VQR, applying the general criteria described in Section 3 and further defined by the GEVs in Section 4. The bibliometric simulation considers the scientific production indexed in the WoS, from all assistant, associate and full professors of Italian universities accepted for the evaluation. For each professor we identify the three publications[9] with the highest score based on the VQR's various criteria. Any conflicts in attributing the publications co-authored by scientists from the same university are resolved by applying an algorithm that assigns the publications in such as a way as to achieve the highest possible overall score for the university in question. For publications directed to peer review or "informed peer review" we adopt certain assumptions:

- For products that were directed to informed peer review under the GEV procedures (the "IR" cases in Table 1 and Table 2), we assign a score of 0.5. These are the products where the values of the two bibliometric indicators are inconsistent (high impact factor with low citations or vice versa). We hypothesize that the reviewers would adopt a judgment with "intermediate" modal value under the scale available.

- For products indexed in the WoS but without impact factor (usually conference proceedings) we assign values of: i) nil, if intended for Biology and Medicine (where the GEVs explicitly notified that they would be given nil score); ii) 0.5 if intended for Information engineering (since in this field conference proceedings are considered to have value); iii) 0.25 in all other cases.

---

[8] These groupings result in very evident distortion when the subject categories show significantly different citation distributions. For example in collapsing Hematology and Rheumatology, publications in Rheumatology are heavily penalized, since on a global basis they are on average cited much less.

All of the above assumptions should be recalled for proper interpretation of the results of the analysis. Clearly our simulation of the VQR cannot completely replicate the real evaluation. The real situation is different: products would have been submitted that were not indexed in the WoS; peer-review evaluations would vary from our assumptions, and the process of product selection has certainly been demonstrated as inefficient (Abramo et al., 2013b). However, the correlation between the universities' VQR rankings and the ones achieved by our simulation are very strong (0.80), with a minimum in Physics (0.68) and a maximum in Medicine (0.91).

The second set of evaluation rankings is produced in exactly the same way but no longer scoring the products on the basis of the GEV criteria. Instead we apply an indicator called Fractional Article Impact Index, which is supported by research published in the literature (presented in detail in Section 5.1). The analysis still refers to the 2004-2010 period but for reasons of significance is limited to the eight hard science UDAs[10]. Table 3 presents the dataset used for the simulations. The evaluation exercise involves 84 universities, employing over 30,000 hard science faculty members that would be required to participate in the evaluation. Together these individuals must submit 85,000 research products.

Beginning from the raw data of the WoS and applying a complex algorithm for reconciliation of the author's affiliation and disambiguation of their precise identity, each publication is attributed to the university scientist or scientists that produced it (D'Angelo et al., 2011). For the 2004-2010 period, the algorithm identifies over 500,000 authorships (last line, column 5), for roughly 244,000 individual publications.

*Table 3: Dataset of the analysis*

| UDA | No. of universities | Research staff subject to evaluation* | Total products to present | Total authorships indexed in the WoS | WoS products selected for bibliometric evaluation† |
|---|---|---|---|---|---|

---
[9] The VQR specified that professors with less seniority would submit only one or two publications. Our model replicates the specific criteria and numbers of submissions.
[10] Mathematics and computer science, Physics, Chemistry, Earth sciences, Biology, Medicine, Agricultural and veterinary sciences, Industrial and information engineering are included. The Civil engineering and architecture UDA is excluded because it includes disciplines that are not hard sciences.

| | | | | | |
|---|---|---|---|---|---|
| Mathematics and computer science | 65 | 3,030 | 8,415 | 27,318 | 6,883 (81.8%) |
| Physics | 62 | 2,096 | 5,864 | 72,312 | 5,542 (94.5%) |
| Chemistry | 58 | 2,741 | 7,591 | 61,306 | 7,375 (97.2%) |
| Earth sciences | 47 | 1,011 | 2,786 | 9,002 | 2,345 (84.2%) |
| Biology | 68 | 4,561 | 12,573 | 64,463 | 11,537 (91.8%) |
| Medicine | 62 | 9,533 | 26,816 | 156,834 | 22,261 (83.0%) |
| Agricultural and veterinary sciences | 56 | 2,866 | 7,958 | 28,284 | 6,219 (78.1%) |
| Industrial and information engineering | 70 | 4,776 | 13,016 | 74,322 | 11,406 (87.6%) |
| Total | 84 | 30,614 | 85,019 | 503,937§ | 73,568 (86.5%) |

*\* i.e. required under VQR to submit at least one product.*
*† In brackets the incidence of products selected compared to products to be presented.*
*§ Eliminating multiple counts for coauthored products, the total is 244,082 individual publications.*

In the two scenarios we are currently considering, the algorithm for the selection of the actual products to be submitted to evaluation then identifies 73,568 publications[11], which is 11,500 less than required under the original VQR criteria. This shortfall is due to 8.5% of the professors not having any publication listed in the WoS, and a further 9.6% who have publications but less than the number that should be submitted.

We observe that the sampling of products to be submitted for evaluation results in the simulation being based on a total of just over a quarter[12] of the WoS-indexed scientific production from the national research staff in the eight UDAs.

---

[11] The true total of individual publications is 65,482: some are represented more than once, being selected for co-authors from different universities.

[12] The ratio is 65,482 individual publications selected divided by the 244,082 indexed in the WoS and attributed to at least one faculty member required to participate in the VQR.

## 5.1 The Fractional Article Impact Index

In this subsection we identify the principle limits of the VQR bibliometric criteria for product evaluation and propose an indicator of value that overcomes these limits. The limitations concern i) the use of the journal impact factor; ii) the failure to consider product quality values as a continuous range, and iii) the full counting of the submitted publications regardless of the number of co-authors.

Each research product embedding production of new knowledge holds a value that is broadly associated with its impact on scientific advancements in the relevant scientific domain. As proxy of impact, bibliometricians typically adopt the number of citations for the researchers' publications. The use of the impact factor (or other similar metrics) of journals where the product is published, has been the object of numerous criticisms by bibliometric experts (Moed and Van Leeuwen, 1996; Seglen, 1997; Glanzel and Moed, 2002; Weingart, 2004). Among other factors, the reliability of citations in representing the true value of a scientific article depends on the so-called citation window, i.e. the time lapse between the publication date and the moment of observing the number of citations received. In fact, Abramo et al. (2010b) have demonstrated that the impact factor can actually be a better predictor of the real impact of an article than the citations, in the case of very short citation windows (less than two years). This holds true especially in Mathematics. Such a consideration may have influenced the VQR GEVs to choose an evaluation mix that includes the impact factor and to assign it still more determining weight for more recent publications (Table 1 and Table 2). However considering that the VQR citation counts were carried out at 31/12/2011, the citation window appears critical only for publications from 2010[13].

A second limit in the VQR concerns the failure to consider product quality values as a continuous range, instead identifying discrete classes (1, 0.8, 0.5, 0), with the resulting introduction of a value cap being a particular concern. Concerning the use of percentile classes in scoring the products, we note that in fact the percentile approach inevitably compresses the differences. For example in class A for citation value, we will observe top publications that contribute radical advances to knowledge, gathering thousands of citations, together with publications at the 80th percentile, with numbers of citations that are two orders of magnitude less[14].

Another question concerns the choice to use percentile standing within global distribution as the evaluation benchmark. The adoption of international benchmarks is correct when the aim of the evaluation is to carry out strategic analysis, i.e. identifying the research fields where a country is weak or strong as compared to others. The outcome of such analysis is not necessarily that funding must be pulled from the weak fields, and in fact policy makers may even choose to invest more heavily in these fields if they consider them strategic. For example, if VQR showed that Italian medical research is less strong than that in physics, then this is a valid and important finding. Italy evidently has to improve the international standing of its medical research. But another point is if research impact based on worldwide normalization would be used for money allocation within Italy, indeed medical research would be in trouble, and would receive much less support than for instance physics. This is of course socially unacceptable. That is why when the aim of the evaluation is selective funding of institutions based on merit, as in the VQR case, we argue that it would be more appropriate to set the benchmark at the national level. In this manner the national assessment will avoid penalizing and removing incentives from groups involved in catch-up research, or in fields where the nation in question has strategic interests but is not currently on the international frontier. To exemplify this discussion in the Italian case, let us assume that the VQR score in medicine of university A is higher than that of University B. University A will then receive more funds than university B. A more detailed analysis shows though that all professors of university A are oncologists, while those of university B are radiologists, and Italian research in oncology is above world average, while in radiology is below. Applying an international benchmark, as VQR did, we can never know to what extent the better performance of oncologists of university A is due to their own capacities or to the heritage in the field transmitted by their predecessors: it may well be that radiologists of university B are much more capable but they are at the early stage of a catch-up research. The adoption of the national benchmark would allow rewarding of the best performers in such cases, by controlling for the world standing in each field. The international benchmark instead invariably leads to allocate fewer resources to all universities active in "nationally weak" fields.

---

[13] In a few disciplines, the citation window may be somewhat critical for publications in the second half of 2009.

[14] For example in the discipline "Genetics & heredity", a simple WoS search for 2008 global publications shows that the top publication has received 2,300 citations, while those in the 80th percentile have received only 22.

In principle, impact would correctly be evaluated by an indicator that rescales each publication's citations in terms of a parameter typical of the domestic reference distribution for that publication. As identified in a previous work of Abramo et al. (2012), this parameter would be the average value of the citation distribution of all national publications cited, from the same year and subject category.

A further critical issue is that the VQR has completely ignored the different contributions of the scientists to the submitted publications in those cases where these are the result of joint work. This situation refers to all coauthored publications. We take the example of two products with exactly the same value but with one authored by a researcher working completely alone and the other by 10 researchers: under the VQR rules, both the "partial" and "full" products score equally in contributing to the university rankings. Potentially exaggerating such effects is the fact that the number of citations that a publication receives is correlated to the number of co-authors that produce it (Harsanyi, 1993; Bridgstock, 1991; Abt, 1984). There are fields of physics, such as high energy physics, with scientists authoring hundreds of publications per year together with hundreds of co-authors. One cannot conclude that they are more productive than a theoretical physicist who publishes a couple of articles per year on his own. Fractional counting, in line with the microeconomic theory of production, is a means to avoid such distortions. It may be objected that the adoption of fractional counting would discourage collaboration behavior. We argue that research collaboration is ever growing, and unavoidable in a number of fields, in particular in physics and medicine. One should also think of the negative effects of straight counting, especially in countries like Italy. Cunning Italian researchers are signing more and more articles that they have never contributed to. Real authors would not care about such "donations", as long as straight counting is adopted. Rates of favoritism in Italy are extraordinary high and opportunistic behavior is the norm rather than the exception (Perotti, 2008). In our view, the choice of bibliometric methodologies should be based also on the distinctive features of the context where they are applied.

For the above reasons, we believe it is more appropriate to fractionalize the bibliometric score for a researcher's publications on the basis of the number of co-authors (Abramo et al., 2013b; Trueba and Guerrero, 2004; Zuckerman, 1968). In the life sciences, the fractionalization can be further refined based on the positioning of the author in the byline[15].

To overcome the weaknesses illustrated in the VQR evaluation, in our second scenario we use the indicator Fractional Article Impact Index (*FAII*) to measure the score of the publications. For any publication *i* authored by researcher *j* the indicator is defined as:

$$FAII_{ij} = \frac{c_i}{\bar{c}_i} f_{ji} \qquad [1]$$

Where:

$c_i$ = citations received by publication *i*;

$\bar{c}_i$ = average citations received by all cited Italian publications of the same year and subject category of publication *i*;

$f_{ji}$ = fractional contribution of the researcher j to publication *i*. Fractional contribution equals the inverse of the number of authors, in those fields where authors appear in alphabetical order in the byline, but assumes different values in the life sciences, where we give different weights to each co-author according to their order in the byline and the character of the co-authorship (intra-mural or extra-mural)[16].

Fractional counting instead of straight counting is not a stringent improvement when applied to three products only per researcher. It is definitely so when applied to overall output of researchers, as done in section 6. Given the concerns over the citation window, noted above, the rating for publications from 2010 is obtained using the same formula [1], but considering the journal impact factor in place of the citations.

### 5.2 Analysis and discussion

The bibliometric simulation of the scenarios is conducted by selecting the best products according to the evaluation criteria for each scenario and then calculating the average score of the products selected. For the first scenario the score derives from the criteria defined by the GEVs. In the second scenario the score derives from the Fractional Article Impact Index.

For each university we calculate the ratio between the total score amounting from the products selected for evaluation and the total number of products to be presented. From these ratios we finally assemble the ranking lists, both at the overall university level and for the national UDAs. For reasons of significance, we exclude the smallest universities from the rankings, as follows:

- those where there are less than 10 research staff members required to submit publications in a particular UDA, for that UDA ranking list;
- those with less than 30 research staff members, for the overall ranking lists.

---

[15] In the life sciences, widespread practice is for the authors to indicate the various contributions to the published research by the positioning of the names in the byline.

[16] If first and last authors belong to the same university, 40% of the contribution is assigned to each of them; the remaining 20% is divided among all other authors. If the first two and last two authors belong to different universities, 30% of the contribution is assigned to first and last authors; 15% is attributed to second and last author but one; the remaining 10% is divided among all others. The weighting values were assigned following advice from senior Italian professors in the life sciences.

Table 4 presents the descriptive statistics stemming from comparison of the two bibliometric simulations. At the overall university level, the two scenarios seem partly overlapping: the correlation index of scores for the two lists is 0.81. However at the UDA level, significant differences emerge: the correlation indexes vary from a minimum of 0.32 in Physics to a maximum of 0.95 for Agricultural and veterinary sciences. In Physics the two scenarios differ notably, with 42 out of 43 universities (97.7%) changing rank. The average variation between the two scenarios is 11.7 positions and the median is 9, with maximum variation at 35. In Physics, particularly in the fields of particle and high-energy physics, research is often conducted through so-called "grand experiments". The results typically have high scientific impact and are accredited to a large part of the research staff of the partner organizations. They are disseminated through publications with hundreds or even thousands of co-authors. Thus the fractionalization of the author contributions in scenario 2 (applying FAII), gives scores that are much different for this type of product when compared to the rankings derived from GEV criteria.

The two scenarios also give notably different ranking lists in Chemistry. Here the score correlation index is only 0.44 and variations of position affect 41 out of 42 universities (97.6%), with average variation of 8 positions and the median at 5. One university shifts a full 36 positions between the two scenarios.

The other UDAs can be clustered in two groups: Biology, Mathematics and computer science, and Industrial and information engineering show score correlation indexes for the two scenarios ranging between 0.64 and 0.85. The average variations in position oscillate between 6.0 (in Industrial and information engineering) and 6.8 (in Biology). The median shifts are between 4 and 5.5 positions, and maximums vary from 26 to 30 positions. The remaining three UDAs (Medicine, Earth sciences, Agricultural and veterinary sciences) show the maximum score correlations between the two scenarios (0.88. 0.91 and 0.95). The convergence between the two scenarios is clear when examining ranking variations: for Medicine the average variation is 3.6, with median at 2; in Agricultural and veterinary sciences (UDA with the least number of universities, at 28) the variations drop to an average of 1.7 with median at 1.0.

*Table 4: Comparison of scores and ranking lists produced from bibliometric simulation of the VQR: a) applying the VQR GEV criteria; b) applying Fractional Article Impact Index*

| UDA | No. of universities | % shifting rank | Average shift | Median shift | Max shift | Score correlation |
|---|---|---|---|---|---|---|
| Mathematics and computer science | 50 | 88.0% | 6.1 | 4 | 26 | 0.85 |
| Physics | 43 | 97.7% | 11.7 | 9 | 35 | 0.32 |
| Chemistry | 42 | 97.6% | 8.0 | 5 | 36 | 0.44 |
| Earth sciences | 30 | 76.7% | 2.9 | 2 | 10 | 0.91 |
| Biology | 50 | 90.0% | 6.8 | 4.5 | 28 | 0.64 |
| Medicine | 43 | 86.0% | 3.6 | 2 | 16 | 0.88 |
| Agricultural and veterinary sciences | 28 | 78.6% | 1.7 | 1 | 7 | 0.95 |
| Industrial and information engineering | 46 | 87.0% | 6.0 | 4.5 | 30 | 0.85 |
| Total | 61 | 96.7% | 5.3 | 3 | 24 | 0.81 |

Since national evaluation exercises often adopt the practice of grouping performance by quartile, we repeat the comparison of the scenarios taking this approach. Table 5 presents the summary of the second analysis. The correlation between the overall ranking lists (based on quartiles) is very high (correlation index 0.87). However a quarter of the 61 universities evaluated change performance quartile between the two scenarios. Three universities actually jump two quartiles[17]. From the last column of the table we see that 12.5% of the top quartile universities under the GEV criteria are no longer "top" under the FAII scenario.

Descending to the UDA level, the comparison between the two scenarios shows much more pronounced differences. For Physics, the Spearman index (0.29) shows an almost total lack of correlation. The average and median quartile variation for the universities is 1. There are actually four cases of three-quartile shifts: three universities in the first quartile under scenario 1 drop to the last scenario under scenario 2, while one university jumps from last to first quartile. In Chemistry there is again a case of a university that places at the top under scenario 1 and in the last quartile in scenario 2, and another university that shifts from last to first. The correlations are higher in the other UDAs, consistent with our previous comparison of scenarios. Still the data from the last column of Table 5 show significant incidence of critical cases, where universities are ranked in the first quartile under scenario 1 but not in scenario 2. The observation is critical, precisely because in many national systems, resources are focused on universities in the first quartile under performance-based funding schemes.

In Physics, six universities out of the 11 in the first quartile for scenario 1 are no longer at the top under scenario 2. The situation for Biology is much the same (7 cases out of 13, or 53.8%). In all the other UDAs, the incidence of such cases is between 25% and 50%.

*Table 5: Comparison of quartile ranking lists from the VQR and FAII simulations*

| UDA | No. of universities | % shifting quartile | Average shift | Median shift | Max shift | Correlat. | From top to non-top |
|---|---|---|---|---|---|---|---|
| Mathematics and computer science | 50 | 34.0% | 0.4 | 0 | 2 | 0.80 | 30.8% |
| Physics | 43 | 65.1% | 1.0 | 1 | 3 | 0.29 | 54.5% |
| Chemistry | 42 | 47.6% | 0.6 | 0 | 3 | 0.61 | 36.4% |
| Earth sciences | 30 | 40.0% | 0.4 | 0 | 1 | 0.85 | 25.0% |
| Biology | 50 | 36.0% | 0.4 | 0 | 2 | 0.81 | 53.8% |
| Medicine | 43 | 30.2% | 0.3 | 0 | 2 | 0.85 | 36.4% |
| Agricultural and veterinary sciences | 28 | 35.7% | 0.4 | 0 | 1 | 0.86 | 28.6% |
| Industrial and information engineering | 46 | 43.5% | 0.5 | 0 | 3 | 0.71 | 41.7% |
| Total | 61 | 23.0% | 0.3 | 0 | 2 | 0.87 | 12.5% |

---

[17] One university drops from the first to third quartile and two jump from the third to first.

## 6. VQR ranking versus productivity ranking

Productivity is the quintessential performance indicator of any production system, therefore the ranking list resulting from a research assessment exercise should primarily reflect the productivity performance of the assessed organizations.

However the "average quality" of a limited sample of products (3 for each scientist in seven years) does not necessarily reflect the "productivity" of the research institutions, i.e. the impact of its overall research product. Therefore we will determine the difference between the VQR rankings ("scenario 1") and the research productivity rankings of institutions, measured as specified in the next section.

### 6.1 Measuring research productivity at institutional level

Different universities have varying research fields and staff per field, and different fields also have varying intensity of publication. Thus to compare the productivity of universities it is necessary to first begin with the measure of productivity in each of their fields. In the Italian university system, each academic is classified in one and only one field, named as their scientific disciplinary sector (SDS).

At research field level (SDS), the yearly average productivity $FSS_S$ over a certain period for researchers in a university in a particular SDS is measured as follows[18]:

$$FSS_S = \frac{1}{w_S} \sum_{i=1}^{N} \frac{c_i}{\bar{c}} f_i \qquad [2]$$

Where:
$w_S$ = total salary of the research staff of the university in the SDS, in the observed period;
$N$ = number of publications of the research staff in the SDS of the university, in the period of observation;
$c_i$ = citations[19] received by publication $i$;
$\bar{c}$ = average citations received by all cited publications of the same year and subject category of publication $i$[18];
$f_i$ = fractional contribution of researchers in the SDS of the university, to publication $i$, calculated as described above.

$FSS_S$ is the basis for measuring research productivity at higher levels of aggregation. Specifically, the productivity $FSS_U$ of a university in a specific UDA $U$, is:

$$FSS_U = \sum_{i=1}^{N} \frac{FSS_{S_i}}{\overline{FSS_{S_i}}} \frac{S_{RS_i}}{S_{RS_u}} \qquad [3]$$

With:
$S_{RS_i}$ = total salary of the research staff of the university in the SDS $i$, in the observed period;
$S_{RS_u}$ = total salary of the research staff of the university in the UDA $U$, in the observed period;
$N$ = number of SDSs of the university in the UDA $U$;
$\overline{FSS_{S_i}}$ = national $FSS_S$ in the SDS $i$.

The formula [3] implies:

---

[18] The theoretical background of the formula may be found in Abramo and D'Angelo (2014).
[19] Or impact factor, for 2010 publications only

- the scaling of the SDS performance to the national average, to take into account the variability of publication and citation intensity among SDSs;
- the weighting of the SDS contribution to the performance of the relevant UDA, to take into account the size of its research staff and, specifically, the incidence of its total salary with respect to the total salary of the UDA.

For measurement of the research productivity of the whole university (*FSS*) the procedure is exactly the same: the only thing that changes is the size weight of the SDS, which is no longer with respect to the other SDSs of the UDA, but rather to all the SDSs of the university. In formula:

$$FSS = \sum_{i=1}^{M} \frac{FSS_{S_i}}{\overline{FSS_{S_i}}} \frac{S_{RS_i}}{S_{RS}} \qquad [4]$$

With:
$S_{RS}$ = total salary of the research staff of the university, in the observed period;
$M$ = number of SDSs of the university;
$\overline{FSS_{S_i}}$ ; $FSS_{S_i}$; $S_{RS_i}$ same as above.

National rankings by $FSS_U$ and $FSS$ at both the UDA and at the university level can be constructed. As specified in the previous analysis, for robustness reasons we excluded:
- Universities with less than 10 scientists working in the UDA, for rankings at UDA level;
- Universities with less than 30 scientists working in the 8 total UDAs, for the global ranking.

**6.2 Analysis and discussion**

Table 6 provides a summary comparison between the scenario 1 rankings (VQR-GEV) and the institutional rankings based on productivity. The Pearson correlation index for the two overall score lists is 0.81. However 56 of the 61 universities evaluated (91.8%) hold different positions in the two rankings, with an average shift of 7.1 positions and median of 4. One of the universities actually shifts 32 positions, and another shifts 40 positions in the rankings for the Mathematics and computer science UDA. In fact at the UDA level, significant differences again emerge. Physics is once more the most problematic discipline. With the score correlation index for the two rankings at 0.37, we can safely affirm that the two evaluations are independent. The average shift of the universities for the two Physics rankings is 11.4 positions and the median is 11. Two universities shift a full 36 positions. In terms of maximum shift, the situation for the Mathematics UDA is again notable. Mathematics has the third-largest average ranking shift (8.6), after Physics and Biology (9.0) and Mathematics is second only to Biology in median shift (5 for Mathematics, 7.5 for Biology). In effect, a full five UDAs can be grouped as showing similar situations: Mathematics and computer science, Biology, Chemistry, Earth sciences and Industrial and information engineering. In this group, the score correlation index for the two rankings is similar, varying from minimum 0.55 to maximum 0.76. Very similar ranges also occur for the central tendency indicators of the distribution of the rankings variations: the median varies between 5 and 7.5 and the mean between 5.9 and 9. In Medicine and Agricultural and veterinary science the correlation between the two lists is clearly greater (Pearson correlation indexes: 0.78 and 0.80). In these UDAs the ranking shifts between the two evaluations are less (averages: 3.8; 5.1; medians: 2.5; 3).

*Table 6: Comparison between scores and ranking lists from the simulated VQR and those based on FSS*

| UDA | No. of universities | % shifting rank | Average shift | Median shift | Max shift | Score correlation |
|---|---|---|---|---|---|---|
| Mathematics and computer science | 50 | 98.0% | 8.6 | 5 | 40 | 0.76 |
| Physics | 43 | 93.0% | 11.4 | 11 | 36 | 0.37 |
| Chemistry | 42 | 97.6% | 6.9 | 5.5 | 23 | 0.74 |
| Earth sciences | 30 | 86.7% | 5.9 | 6 | 18 | 0.55 |
| Biology | 50 | 94.0% | 9.0 | 7.5 | 34 | 0.70 |
| Medicine | 43 | 86.1% | 5.1 | 3 | 15 | 0.80 |
| Agricultural and veterinary sciences | 28 | 82.1% | 3.8 | 2.5 | 13 | 0.78 |
| Industrial and information engineering | 46 | 93.5% | 8.3 | 6 | 34 | 0.63 |
| Total | 61 | 91.8% | 7.1 | 4 | 32 | 0.81 |

As we did for the previous analysis, we repeat the comparison of the scenarios with the subdivision of the rankings in quartiles (Table 7). The correlation index between the two overall quartile rankings is 0.77 with an average quartile variation of 0.4 and median of nil. The maximum shift of two quartiles occurs for five universities: two move from the first quartile under the VQR simulation to third for productivity and three other universities make the same shift but in reverse. The rankings lists constructed for the individual UDAs are much less superimposable. As usual, Physics shows practically nil correlation between the rankings (0.25): on average, the 43 universities active in the discipline shift one quartile between the two rankings. For four more UDAs the correlations appear weak: the Spearman index is less than 0.6 in Earth sciences, Biology, Industrial and information engineering and Mathematics and computer science.

In two of these UDAs the median quartile variation is nil but there are individual cases of shifts of three quartiles. In Mathematics and computer science, a university drops from the first quartile under the VQR ranking to last quartile for productivity. Chemistry, Medicine and Agricultural and veterinary sciences all show two or three cases of universities leaping ahead the full three quartiles possible. In the last column of Table 7, we see that more or less half of the top universities under the VQR rankings are no longer top by productivity. In Biology, only four of the 13 top VQR universities remain top by productivity: four of them drop two quartiles, and one falls to the last quartile.

*Table 7: Comparison of VQR quartile ranking lists and FSS ranking lists for institutional productivity*

| UDA | No. of universities | % shifting quartile | Average shift | Median shift | Max shift | Correlat. | From top to non-top |
|---|---|---|---|---|---|---|---|
| Mathematics and computer science | 50 | 46.0% | 0.6 | 0 | 3 | 0.60 | 46.2% |
| Physics | 43 | 60.5% | 1.0 | 1 | 3 | 0.25 | 38.5% |
| Chemistry | 42 | 59.5% | 0.7 | 1 | 2 | 0.69 | 45.5% |
| Earth sciences | 30 | 60.0% | 0.8 | 1 | 3 | 0.52 | 37.5% |
| Biology | 50 | 52.0% | 0.7 | 1 | 3 | 0.60 | 69.2% |
| Medicine | 43 | 48.8% | 0.6 | 0 | 2 | 0.73 | 45.5% |
| Agricultural and veterinary sciences | 28 | 46.4% | 0.5 | 0 | 2 | 0.77 | 42.9% |
| Industrial and information | 46 | 47.8% | 0.7 | 0 | 3 | 0.56 | 50.0% |

| | engineering | | | | | | | |
| | Total | 61 | 34.4% | 0.4 | 0 | 2 | 0.77 | 18.8% |

## 7. Conclusions

National research assessment exercises are becoming ever more common. The intended aims consistently concern support for various objectives in national research policy. Many of the countries implementing these exercises are characterized by non-competitive higher education systems, where evaluation was virtually unknown until a few years ago.

Unfortunately, all that glitters is not gold. The wide diversity in practices chosen by the different national bodies in charge of evaluation, even in the simple issue of selecting time frames, confirms that we are still very much in a trial and error phase in terms of methodologies. In fact, bibliometric evaluation is still in rapid evolution and as yet has few consolidated standards.

Scholars in the field must thus offer clear contributions, defining the methodological requirements for evaluation exercises so that they can be planned to meet the latest standards, for results that are coherent with the intended macroeconomic objectives. The specialists must then also contribute to the post-facto verification of the outcomes.

In this work we have concentrated on the issue of post-facto verification. Applying bibliometric criteria, we have simulated the VQR national evaluation exercise recently completed in Italy. Taking the simulated ranking results of the VQR, we then compared them to two other potential evaluation scenarios.

Our analyses show that that there are substantial weaknesses in the VQR methodology and scoring criteria as applied for evaluation of products in the hard sciences. At the theoretical level, the criteria seem planned without support from the evaluation literature. In their actual application they then create critical situations, particularly in certain disciplines. In particular, the decision to ignore co-authorship of the products has probably resulted in rankings that are rather different from those obtained with the methodology employed here. This situation is remarkably exaggerated in the Physics discipline. Equally preoccupying is the choice to group the scores for product quality in four classes, with a value cap that prevents detection of excellence and blocks its relative contributions to institutional scores. These particular choices are clearly doubtful at the theoretical level. For the Italian case, at this point we can only hope that in such large-scale analysis the impact on the institutional rankings will tend to balance out.

But the authors' greatest concern is for the choice to base evaluation on a subpopulation of the universities' total research output – a planning decision which is also made in other countries. For assessment exercises to meet the aims of stimulating increased performance, the organizations must be evaluated on the basis of their real productivity, not the average quality of a limited set of their products. The productivity indicator proposed, FSS, embeds both quantity and quality of production and thus will not induce distorted behavior in scientists, who have varying and valid motives for selecting different frequencies and styles of publication. Furthermore, adopting fractional counting rather than full counting helps to avoid possible opportunistic behavior by scientists, who could increase their outputs by signing works for which they offered no real contribution. In fact if the full counting method is applied then the true authors would experience no penalty from this behavior, whereas fractional counting obviously results in negative implications and would likely lead them to decline such involvement. The necessity of examining productivity is even more acute under a bibliometric approach, such as in the VQR. Our simulations show that roughly half of the universities that classified at the top of the national UDAs under VQR criteria would not rate at the top on the basis of their productivity. There are numerous cases of VQR top universities that actually end up at the bottom under rankings for productivity, or that leap from bottom to top. The scope of the ranking problems detected is net of further distortions caused by the inefficient selection of the limited set of products, which is also known to occur. Given the financial compensation based on the university ratings, the research raises serious questions about the VQR's true socio-economic impact. The result of this or similar evaluation exercises will clearly be distorted information about the relative quality of many universities, with resulting unjust allocation of resources. Thus the Italian case indicates the need for the various national agencies to take more care in the planning stages of assessment exercises, and to call on assistance from the most qualified professionals in the field. We must naturally accept that there will be a margin of error in any measurement system, however it is much more difficult to accept distortions due to use of measurement systems not in keeping with progress in theory of research evaluation.